\documentclass[conference,10pt]{IEEEtran}
\usepackage[dvips]{color}
\usepackage{epsf}
\usepackage{times}
\usepackage{epsfig}
\usepackage{graphicx}
\usepackage{tabularx}
\usepackage{amsmath}
\usepackage{amssymb}
\usepackage{amsxtra}
\usepackage{amsthm}
\usepackage{bbm}

\usepackage{here}
\usepackage{rawfonts}
\usepackage{times}
\usepackage{url}
\usepackage{cite}
\usepackage{comment}
\usepackage[utf8]{inputenc}
\usepackage{caption}
\usepackage{subcaption}

\usepackage{pstricks}
\usepackage{algorithm}
\usepackage{algpseudocode}
\usepackage{lipsum}
\usepackage{mathtools}

\usepackage{geometry}
\makeatletter 

\IEEEoverridecommandlockouts

\begin{document}
\title{\huge Latency Optimization in LEO Satellite Communications with Hybrid Beam Pattern and Interference Control} 
\author{ 
	\IEEEauthorblockN{ Qianqian Zhang$^1$, Ye Hu$^2$, and Minchae Jung$^3$  }
	
	\IEEEauthorblockA{\small 
		$^1$Department of Electrical and Computer Engineering, Rowan University, NJ, USA,
		Email: \url{zhangqia@rowan.edu}\\
		$^2$Department of Industrial and Systems Engineering, University of Miami, FL, USA, 
		Email: \url{yehu@miami.edu}\\
		$^3$Department of Electronics and Information Engineering,
		Sejong University, Seoul, South Korea, 
		Email: \url{mcjung@sejong.ac.kr}
	}
} 
\maketitle
\vspace{-0.2cm}

\setlength{\columnsep}{0.55cm}
\begin{abstract} 
The rapid advancement of low Earth orbit (LEO) satellite communication systems has significantly enhanced global connectivity, offering high-capacity, low-latency services crucial for next-generation applications. 
However, the dense configuration of LEO constellations poses challenges in resource allocation optimization and interference management, complicating coexistence with other communication systems.  
To address these limitations, this paper proposes a novel framework for  optimizing the beam scheduling and resource allocation in multi-beam LEO systems.    
To satisfy the uneven terrestrial traffic demand, a hybrid beam pattern is employed to enhance the downlink quality of service and minimize the transmission latency from LEO satellites to ground user terminals. 
Additionally, a dynamic co-channel interference (CCI) control mechanism is developed to mitigate inter-beam interference within the LEO constellation and limit cross-system interference affecting protected users from other networks.  
The problem of user-beam-frequency allocation with power optimization is formulated as a mixed-integer dynamic programming model and solved using a low-complexity neural network-based graph generation algorithm. 
Simulation results show that the proposed approach outperforms the baseline methods of full frequency reuse and single-channel transmission, and highlights  the potential for further performance improvement with multi-user transmissions.
 
\end{abstract}

\IEEEpeerreviewmaketitle

\section{Introduction}

Low Earth orbit (LEO) satellite systems have attracted increasing attention due to the continued  deployment of mega-constellations, such as Starlink  and Oneweb \cite{deng2021ultra}.   
With hundreds to thousands of satellites in orbit, each equipped with multiple antennas supporting high-gain beams,  LEO constellations can efficiently deliver seamless and global coverage with high-capacity communication service.  
Recent advancements in satellite technology with decreased launch costs enable  LEO constellations as a cost-effective and scalable solution for extending broadband internet access to underserved and remote regions, as well as complementing  existing terrestrial networks with enhanced coverage, resilience, and capacity \cite{kim2023downlink}. 

On the user side, as demand for real-time applications like video conferencing and autonomous systems continues to grow,  low-latency and high-throughput communication becomes increasingly  critical.  
Despite their advantages, LEO satellite systems still experience a higher round-trip latency (typically in tens of milliseconds), compared to ground-based networks based on optical fiber.     
However, since the signal propagation speed in free space is approximately  $47\%$ faster than in fiber-optic cables, LEO satellites have the theoretical potential  to achieve a lower latency in long-distance communications \cite{handley2018delay}. 
Therefore, there exists a clear opportunity for technological innovations to close the performance gap and optimize LEO constellation systems for latency-sensitive applications. 

The proliferation of LEO satellite systems also introduces a major challenge due to the increased interference from dense satellite deployments. 
Under International Telecommunication Union (ITU) regulations, LEO satellites must avoid interference with geostationary (GEO) networks by maintaining the equivalent power flux density (EPFD) within specified limits, which necessitates frequent beam adjustments or band-switching to prevent disruption \cite{su2019broadband}.   
Beyond GEO interference, LEO systems can also impact radio telescopes and astronomical systems that rely on detecting faint signals.  
These passive users are highly sensitive to overlapping frequencies or harmonics, despite certain frequency bands being dedicated for radio astronomy. 
Furthermore, ground cellular networks face similar issues, as  LEO satellites can operate in overlapping frequency bands, and the growing interest in integrated terrestrial-space communication systems complicates the interference landscape \cite{handley2018delay}. 
To address these challenges, dynamic spectrum management and real-time beam control are essential to reduce interference and support harmonious coexistence among   communication networks.

Various aspects of  interference control and performance optimization for LEO communications have been explored in \cite{kim2023downlink} and \cite{liu2020dynamic, lin2024satellite, cui2022latency, chu2020robust, zheng2024traffic}.   
The authors in \cite{kim2023downlink} analyze the performance of multi-beam satellite communications by characterizing the received powers of both desired and interference signals, and \cite{liu2020dynamic} proposed a  beam shut-off algorithm to avoid co-channel interference (CCI) between multiple satellites.  
To efficiently allocate communication resources, \cite{lin2024satellite} focused on the beam hopping scheduling to meet uneven terrestrial traffic demands,  
while \cite{chu2020robust} examined  the non-orthogonal multiple access scheme to support a large number of ground devices distributed over a large area.
However, most existing works considered the downlink data rate or transmit power as prime performance metrics, neglecting the importance of latency in the quality of service (QoS).   
Although \cite{cui2022latency} and \cite{zheng2024traffic} jointly considered the resource allocation and latency optimization for LEO satellites, the coexistence challenge of LEO systems with other communication networks is not addressed in their beam pattern designs.

This paper aims to optimize the resource allocation and beam scheduling in multi-beam LEO satellite systems with dynamic interference control.  
To support efficient transmissions, a hybrid pattern combining a wide beam with multiple spot beams  is employed to minimize the downlink latency from each LEO satellite to  ground user terminals (UTs). 
Additionally, dynamic CCI control not only considers inter-beam interference within LEO constellations, but also evaluates the impact on protected users from other communication systems. 
To address the mixed integer dynamic programming problem for the latency optimization, we decompose the task into two steps: beam-UT association and  beam-channel allocation.
Then, a graph generation algorithm based on neural networks is proposed to find the optimal resource allocation scheme with low-computational complexity and minimize the expected latency of LEO downlink communications. 
Simulation results show that the proposed approach outperforms other reference schemes.









\section{System Model and Problem Formulation} 

\begin{figure}[!t]\vspace{-0.3 cm} 
	\begin{center}  
		\includegraphics[width=6.8cm]{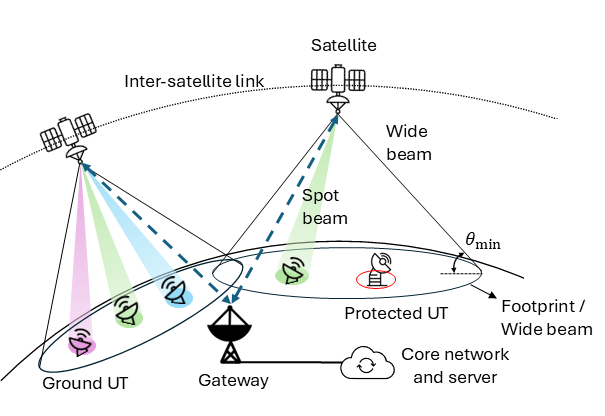} \vspace{-0.3 cm} 
		\caption{\label{beam}\small Hybrid beam coverage in LEO constellation system. }  
	\end{center}\vspace{-0.7 cm}  
\end{figure}

Consider a LEO satellite  system that provides downlink transmission services to ground UTs, using a  multiple-frequency time-division multiple access (TDMA) communication model with a hybrid beam pattern. 
In particular, the location of satellite $s$ is denoted as $\boldsymbol{s}=(x,y,h)$ under the Earth-Centered Inertial  coordinate framework \cite{deng2021ultra}, and its service area, also called footprint,  is determined by the  minimum elevation angle $\theta_{min}$, as shown in Fig. \ref{beam}.  

For seamless coverage, a LEO constellation is densely deployed so that each UT is covered by at least one satellite. When activated, a UT connects to a satellite that locates with UT's minimum elevation angle and ensures a long service duration. 
After the connection establishment, the UT sends requests to the satellite, which forwards them to a terrestrial gateway linked with core network and servers. 
After processing, the response is sent back through the forward link to the satellite and then to the UT. This study focuses on the downlink transmission of the forward link from LEO satellites to ground UTs.

\subsection{Communication Model}

For downlink transmissions, each satellite is equipped with multiple phased array antennas to employ a hybrid beam pattern \cite{su2019broadband}, where a fixed wide beam covers the entire service area for control signaling, while multiple spot beams are steered towards active ground UTs to deliver higher power  for increased data rates and  flexible on-demand service. 
As shown in Fig. \ref{beam}, each spot beam covers a subarea of the satellite's footprint. 
All LEO satellites reuse the same downlink bandwidth, while each satellite $s$ divides its bandwidth into $M$ channels, which are reused by $B$ spot beams, where $B>M$ \cite{del2019technical}.   
The direction vector of each beam $b \in \mathcal{B}$ is denoted by $\boldsymbol{b}$ with $|\boldsymbol{b}|=1$. 
  
Consider a downlink channel  from a satellite $s$ and a UT $u$, the channel gain over a spot beam $b$ using a frequency channel $m$ can be expressed as
\begin{equation}\vspace{-0.2cm}
	H_{u,s,b,m} = \frac{G^T_{u,s,b} G^R_{u,s}}{ \xi_{u,s,m} L_{u,s,m} } 
\end{equation}
where $G^T_{u,s,b}$ is the transmitter antenna gain at the satellite, $G^R_{u,s}$ is the receiver antenna gain at the UT,
$\xi_{u,s,m}$ represents the atmosphere attenuation factor following a double log-norm distribution \cite{kim2024performance}, 
and $L_{u,s,m} = ({4\pi d_{u,s} f_m}/{c})^2$ is the free space path loss, 
with $f_m$ as the carrier frequency, $c$ as the speed of light, and $d_{u,s}$ as the UT-satellite line-of-sight distance.    
The off-axis angle between a satellite-UT link and its beam is $\phi = \arccos(\frac{\boldsymbol{b} \cdot (\boldsymbol{u} - \boldsymbol{s})}{|\boldsymbol{b}||\boldsymbol{u} - \boldsymbol{s}|})$, where $\boldsymbol{u}$ is the UT's location. Then, the  antenna gain at the transmitter can be given by \cite{ITURS15283} 
\begin{equation*}\vspace{-0.1cm}
	G^T_{u,s,b}(\phi) = \begin{cases}
		G^T_{max} & \phi \le \phi^{h}\\
		G^T_{max} - 3(\frac{\phi}{\phi^{h}})^2 &  \phi^{h} < \phi \le  \frac{3}{2}\phi^{h} \\
		G^T(\frac{3}{2}\phi^{h}) - 25 \log_{10}(\frac{2\phi}{3\phi^{h}}) &  \frac{3}{2}\phi^{h} < \phi \le \phi^{max} \\
		L^T_F & \phi > \phi^{max}, 
	\end{cases} 	
\end{equation*}
where  $\phi^{h}$ is one half of $3$-dB beamwidth, $G^T_{max}$ is the mainlobe antenna gain, and $\phi^{max}$ denotes the outer-edge of the sidelobe, with $L^T_F = 5$ dBi as the far-out sidelobe level. 
Besides, the  radiation pattern at the UT's antenna can be expressed as \cite{ITURS4656}
\begin{equation*}\vspace{-0.1cm}
		G^R_{u,s,b}(\psi) = \begin{cases}
			G^R_{max}, & \psi \le \psi^{e}\\
			G^R(\psi^{e}) - 25 \log_{10} (\frac{\psi}{\psi^{e}}) & \psi^{e} <\psi \le \psi^{max} \\
			L^R_F & \psi > \psi^{max}, 
		\end{cases}
\end{equation*}
where $\psi$ is the off-axis angle between the UT-satellite link and the receiver's mainlobe, $\psi^{e}$ is the angle separating main and side lobes, $\psi^{max}$ denotes the outer-edge of the sidelobe, and $L^R_F = -5$ dBi. 
Here, we assume that the UT's mainlobe and the associated satellite's beam is perfectly aligned, i.e., the off-axis angles $\phi_{u,s,b} = \psi_{u,s,b} = 0$ if satellite $s$ assigns beam $b$ for downlink transmission towards UT $u$.

Due to satellite antenna imperfections with $L^T_F>0$, CCI occurs between different beams of the same satellite and between neighboring LEO satellites. 
Let $\mathcal{L}_s$ denote the set of satellites with overlapping footprints of satellite $s$, and for simplicity, we assume inter-satellite CCI only occurs within $\mathcal{L}_s$.  
During each time slot $t$, let $P^t_s$ denote the available downlink power for satellite $s$, and $P^t_{s,b}$ be the transmit power allocated to spot beam $b$, where $\sum_{b=1}^{B} P^t_{s,b} \le P^t_s$.  
The downlink signal-to-interference-and-noise ratio (SINR) for UT $u$ served by beam $b$ of satellite $s$ is then
\begin{equation}\vspace{-0.2cm}
	\Gamma^t_{u,s,b,m} =  \frac{ P^t_{s,b,m} |H_{u,s,b,m}|^2}{\sigma^2 + I^t_{u,s,b,m} + O^t_{u,s,m}}  ,
\end{equation}
where 
$P^t_{s,b,m}$ is the power allocated to beam $b$ over channel $m$ with  $\sum_{m=1}^M P^t_{s,b,m} = P^t_{s,b}$. 
Note that, if beam $b$ is allocated with frequency band $m$ for downlink transmission, then $ P^t_{s,b,m}>0$; otherwise, $ P^t_{s,b,m}=0$. 
Here, $\sigma^2$ represents  the Gaussian noise power,  $I^t_{u,s,m}$ and  $O^t_{u,s,m}$ denote the intra-  and inter-satellite CCI power, respectively.  
In particular, 
\begin{equation*}\vspace{-0.1cm}
	I^t_{u,s,b,m} = \sum_{b^{'}\in\mathcal{B},b^{'}\ne b} P^t_{s,b^{'},m} |H_{u,s,b^{'},m}|^2 .  
\end{equation*}
And the inter-satellite CCI can be given by
\begin{equation*}
	O^t_{u,s,m} = \sum_{s^{'} \in \mathcal{L}_s} \sum_{b^{'} \in \mathcal{B}_{s^{'}}}  P^t_{s^{'},b^{'},m} |H_{u,s^{'},b^{'},m}|^2 .
\end{equation*} 
For analysis tractability, we assume that inter-satellite links (ISLs) exist between adjacent satellites to coordinate the co-channel beam planning and UT handover.  Thus, the average value of the inter-satellite CCI  can be estimated.     
Consequently, the downlink  rate from satellite $s$ to UT $u$ can be given as
\begin{equation}\label{equ_rate}
R^t_{u,s} = \sum_{b=1}^{B} \sum_{m=1}^M \mathbbm{1}({c^t_{u,s,b}=1})\cdot \beta \cdot\log_{2} ( 1 + \Gamma^t_{u,s,b,m}),  
\end{equation}
where  $\beta$ is the bandwidth of each frequency channel, $\mathbbm{1}(\cdot)$ is the indication function that outputs $1$ for true statement and $0$ for false statement, and $c^t_{u,s,b} = 1$ indicates that beam $b$ is allocated to UT $u$ in time slot $t$; otherwise, $c^t_{u,s,b} = 0$. 
Here, only when $\sum_{b=1}^{B} \sum_{m=1}^{M} P^t_{s,b,m}\cdot c^t_{u,s,b} > 0$, then UT $u$ is scheduled for downlink transmissions  during time slot $t$.

\subsection{Interference Control} 

LEO constellations can interfere with other systems, like GEO transmissions, radio astronomy, and terrestrial cellular networks using overlapping service areas and  frequencies. To ensure coexistence, dynamic spectrum management is crucial to support real-time beam adjustments and resource allocation based on network conditions and user demands. 
Following ITU's interference regulation \cite{ITURRA2020}, we define  $\mathcal{R}_s$ as a set of protected UTs within satellite $s$'s footprint.    
The EPFD from beam $b$ of satellite $s$ to each protected UT $r \in \mathcal{R}_s$ over frequency channel $m$ is given as
\begin{equation}\label{erfd}
	E^t_{r,s,b,m} =  \mathbbm{1}(a^t_{r,m}=1) P^t_{s,b,m} \frac{G^T_{r,s,b}}{4\pi d_{r,s}^2} \frac{G^R_{r,s}}{G^R_{r,max}} , 
\end{equation}
where $a^t_{r,m}=1$ indicates the protected UT $r$ employed channel $m$ in time slot $t$; otherwise, $a^t_{r,m}=0$. Here,  $d_{r,s}$ is the distance between satellite $s$ and  UT $r$, ${G^R_{r,s}}$ is the receiver's antenna gain of $r$ with respect to satellite $s$, and ${G^R_{r,max}}$ is the maximum antenna gain of  UT $r$. 
Here, we assume the location and antenna pattern of each protected UT are known by each LEO satellite via the network management center, and the EPFD must not exceed the threshold, i.e., $E^t_{r,s} =\sum_{b=1}^{B}  \sum_{m=1}^{M} E^t_{r,s,b,m} \le \kappa$. 

\subsection{Traffic and Latency Model}


Based on the interference control and  communication model, the downlink transmission towards  ground UTs needs to be scheduled to guarantee a good quality-of-service. 
We consider each satellite with a limited-size buffer, and downlink traffic follows the first-in-first-out model, 
subject to a time-to-live (TTL) constraint,  i.e., data is discarded if it is not transmitted within a specified number of time slots. 
Let $Q^t_{u,s}$ represent the amount of data remaining in the buffer of satellite $s$ for UT $u$ at the beginning of time slot $t$.  
The amount of data transmitted during  time slot $t$ is then given as \cite{cui2022latency} 
\begin{equation}
	\Delta Q^t_{u,s} = \min \{ Q^t_{u,s}, R^t_{u,s }  \Delta T \}, 
\end{equation}
where $\Delta T$ is the length of each time slot. Then the amount of remaining data that needs to be transmitted at time slot $t+1$ will be 
\begin{equation}\label{equ_q}
	Q^{t+1}_{u,s} = Q^t_{u,s} - \Delta Q^t_{u,s}. 
\end{equation} 
If a satellite  moves beyond the UT's view, 
a soft handover occurs between satellites via ISLs, and any remaining transmission is then transferred to the next associated satellite. 

Let $\tau^t_{u,s}$ be the duration that UT's downlink data has  waited in satellite's buffer at the beginning of time slot $t$. 
When the data first arrives, the waiting clock is initialized as $\tau^{t^{0}}_{u,s}=0$.   
If UT $u$ is not scheduled or its downlink transmission is incomplete,  $\tau^t_{u,s}$  increments by one time slot $\Delta T$ until all data in the buffer is transmitted, i.e.,   
\begin{equation}
	\tau^t_{u,s} = \begin{cases} 
		\tau^{t-1}_{u,s} + \Delta T, & Q^t_{u,s} \ne 0 \\
		\tau^{t-1}_{u,s},  & Q^t_{u,s} = 0,
	\end{cases} 
\end{equation} 
Then the overall latency that downlink data of UT $u$ waits in the buffer of satellite $s$ will be
	$\tau_{u,s} = t^{*} - t^{o}$, 
where $t^{*}$ is the time slot when transmission is completed, indicated by $Q^{t^{*}}_{u,s} = 0$. 


\subsection{Problem Formulation}
In  dense LEO constellation systems, each satellite must schedule its beam patterns to provide downlink transmission service to multiple ground UTs. 
Specifically, the interference control mechanism is essential  to mitigate the CCI affecting protect UTs, while minimizing the downlink latency by efficiently  allocating power, frequency channels, beams, and time slots for each active UT.   
Considering a time period of $\mathcal{T}$, the optimization problem can be expressed as
\begin{subequations}\label{equsOpt}
	\begin{align}
		\min_{c^t_{u,s,b},P^t_{s,b,m}}  & \tau_s= \sum_{u=1}^{{U}_s} \tau_{u,s} & \label{equOpt}\\
		\textrm{s. t.} \quad 
		& \sum_{b=1}^{B} \sum_{m=1}^M P^t_{s,b,m} \le P^t_s, &\forall t \in \mathcal{T}  \label{con_power_sum} \\ 
		&  P^t_{s,b,m} \ge 0, & \forall b \in \mathcal{B}, \forall t \in \mathcal{T} \label{con_power_each}\\
		&  E^t_{r,s} \le \kappa, & \forall r \in \mathcal{R}_s, \forall t \in \mathcal{T} \label{con_epfd}\\
		& \sum_{m=1}^{M} \mathbbm{1}(P^t_{s,b,m}>0) \le N, &\forall b \in \mathcal{B}, \forall t \in \mathcal{T} \label{con_freq}\\	
		& \sum_{u=1}^{{U}_s} c^t_{u,s,b}  \le 1, &\forall b \in \mathcal{B}, \forall t \in \mathcal{T}  \label{con_beam}\\
		& c^t_{u,s,b} \in \{0,1\}, &\forall u,b,t \label{con_beamchannel}
	\end{align} 
\end{subequations} 
The objective function (\ref{equOpt}) minimizes the total downlink latency for all UTs. 
Constraints (\ref{con_power_sum}) and (\ref{con_power_each}) ensure that each beam is allocated with a transmit power across its associated frequency bands, and the total power is limited by the satellite's energy generation. 
Constraint (\ref{con_epfd}) ensures compliance with EPFD regulations, while (\ref{con_freq}) limits each spot beam to a maximum of  $N$ frequency bands for channel aggregation  during any time interval.   
(\ref{con_beam}) requires that each beam  serves only one UT per time slot\footnote{Advance multiplexing techniques, such as DVB-S2, will be explored in our future work, to serve the UT cluster, instead of an individual UT per beam.}. 
And constraint (\ref{con_beamchannel}) represents the UT-beam allocation.   
Notably, frequency bands can be reused across  different beams if needed. 

\section{Optimization of Downlink Transmissions}

\begin{figure}[!t]\vspace{-0.3 cm} 
	\begin{center}  
		\includegraphics[width=6.8cm]{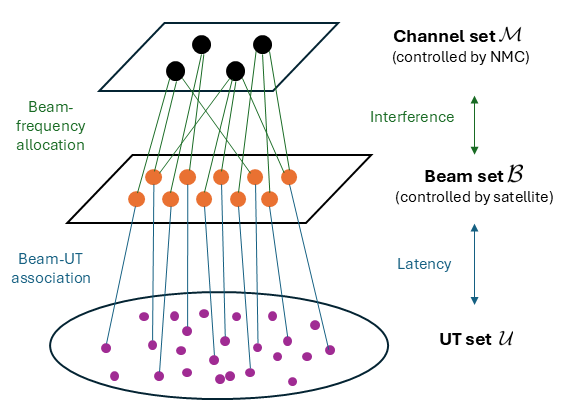} \vspace{-0.2 cm} 
		\caption{\label{node}\small The resource allocation between ground users, LEO satellite, and the NMC panels. Each blue line represents a beam-UT association, and each green link denotes a bean-frequency allocation.}  
	\end{center}\vspace{-0.7cm}  
\end{figure}

The optimization problem in (\ref{equsOpt}) is a non-convex mixed integer problem, which is challenging to find an optimal solution. 
As shown in Fig. \ref{node}, in this section, we will decompose (\ref{equsOpt}) into two subproblems: the UT-beam association and the beam-frequency allocation, with a dynamic power optimization. 


\subsection{Beam-UT Association}

Based on (\ref{con_beam}) and (\ref{con_beamchannel}), for each satellite $s$, the number of active UTs receiving downlink transmissions per time slot will  be equal to the number of spot beams.  
Assuming perfect beam alignment between each UT and its associated beam, the beam index becomes irrelevant \cite{kim2023downlink}. That is, the satellite-UT channel remains unchanged regardless of the assigned beam,  provided that the same frequency band is used, i.e., $H_{u,s,b,m} = H_{u,s,b^{'},m}$. 
Thus, the beam-UT association simplifies to selecting a subset of $B$ UTs from $\mathcal{U}$, following two  criteria: maximizing the spatial separation to allow frequency band reuse among multiple UTs without significant intra-satellite CCI, and minimizing the downlink latency to ensure fast data delivery. 

To achieve both objectives, a UT clustering algorithm based on weighted K-means is employed. 
First, we choose $B$ UTs with the highest latency values $\tau^t_{u,s}$, and use their locations as the initial centers.   
Each UT is then assigned to the nearest cluster based on a distance metric. 
The cluster centers are updated by calculating the weighted average of UTs' locations within each cluster, using the latency values $\tau^t_{u,s}$ as a weight to determine each UT's contribution. 
After convergence, the UT with the highest latency in each cluster is chosen as the service receiver$^1$, i.e., $c^t_{u,s,b}=1$. 
Finally, the transmit power $P_{s,b}$ will be allocated to each selected UT (via its associate beam) in proportion to its remaining downlink data amount $Q^t_{u,s}$.  

\subsection{Beam-frequency Allocation}

With inter-satellite CCI and EPFD requirements, each satellite will rely on the network management center (NMC)  for external information. 
Assuming the NMC controls data of each frequency channel $m\in\mathcal{M}$, real-time data exchange between each satellite and the NMC is necessary for beam-frequency allocation and transmit power optimization. 

\subsubsection{Graph-based model}

Given the beam-UT association, we model the frequency allocation problem as a graph $\mathcal{G}$$=$$\{\mathcal{V},\mathcal{E}\}$, where the node set $\mathcal{V}$$=$$\mathcal{B}$$\cup$$\mathcal{M}$ consists of spot beams and frequency channels. 
Each edge $e_{m,b}$$\in$$\mathcal{E}$ only exists between beam  and channel subsets, representing the allocation of channel $m$ to beam $b$.    
Each beam has a feature vector $\boldsymbol{n}_{b}$ $=$ $[\boldsymbol{u}(b),$ $P_{s,b},Q^t_{u(b),s},$$\tau^t_{u(b),s}]$, where $\boldsymbol{u}(b)$ is the location of the associated UT for beam $b$.  
The feature vector of each frequency node $\boldsymbol{n}_{m}= \boldsymbol{r}_{s,m} $ encodes the location of EPFD-protected UTs that utilize channel $m$ within the footprint of satellite $s$. 


From the perspective of information propagation \cite{li2018learning}, given any edge $e_{m,b}$$\in$$\mathcal{E}$, 
channel $m$ can provide beam $b$ with downlink channel information in a message vector $\boldsymbol{g}^0_{m,b}=$ $[\xi_{u(b),s,m}, $$ L_{u(b),s,m}]$. 
In return, beam $b$ will initialize its power allocation 	$P_{s,b,m} = f_p(\boldsymbol{n}_{b},\boldsymbol{g}^0_{m,b}) $, and then, share the  power and antenna information with channel $m$, via the message vector $\boldsymbol{g}^0_{b,m} = [\boldsymbol{u}_b,P_{s,b,m},	G^T_{u,s,b}, G^R_{u,s,b}]$. 
After receiving the message from all beams, channel node $m$ can aggregate the downlink information, calculate the inter- and intra-satellite CCI, and propagate the information back to each beam $b$ via $\boldsymbol{g}^1_{m,b} = [I_{u(b),s,b,m},O_{u(b),s,b,m}]$. 
Then, beam $b$ will update its power  	$P_{s,b,m} = f_p(\boldsymbol{n}_{b},\boldsymbol{g}^{1}_{m,b}) \triangleq \boldsymbol{g}^1_{b,m}$  to ensure compliance with EPFD regulations and optimize the downlink performance over multiple frequency bands towards its UT. 
In this case, the optimization problem simplifies to the maximization of sum data rates under constraints (\ref{con_power_sum})-(\ref{con_epfd}), which can be solved by  Lagrange dual decomposition, after multiple rounds of information propagation. 
Next, we will study how to allocate each frequency band to satellite beams, i.e., how to create edges between $\mathcal{B}$ and $\mathcal{M}$, using a graph generation approach \cite{liao2019efficient}.

\subsubsection{Generative graph for beam-frequency allocations}

Given the beam-UT association,  $\mathcal{G}$ is initialized with $B$ beam nodes and $M$ channel nodes, but no edge. 
Given independent frequency bands, $\mathcal{G}$ can be constructed  over $M$ sequential steps, and at each step $m=1,\cdots,M$, a subset of beam nodes is selected to allocate the $m$-th frequency channel. 
Notably, adding one more edge $e_{b^{'},m}$ to $\mathcal{G}$ introduces interference to all the other beams $b$ in the  subset with $e_{b,m} \in \mathcal{E}$. 
Therefore, each step of frequency allocation involves two key decisions: whether to add a new edge and, if so, which beam to connect to the $m$-th  channel. 
Then, a graph generative model defines a distribution over the sequence of graph generating decisions over all possible outcomes, using two following equations \cite{li2018learning}:


$f_{addedge}(\mathcal{G},m)=\sigma[f_{ae}(m,\boldsymbol{n}_{\mathcal{E}},\boldsymbol{g}_{\mathcal{E}})]$  takes the existing graph $\mathcal{G}$ as an input, and outputs the   decision on whether to terminate the algorithm or add another edge with channel $m$.   
First, $f_{ae}(m,\boldsymbol{n}_{\mathcal{E}},\boldsymbol{g}_{\mathcal{E}})$  uses the channel feature $\boldsymbol{n}_m$ and message vectors $\boldsymbol{g}_{\mathcal{E}}$ of all existing edges  to compute an attention weight \cite{velickovic2018graph} for channel node $m$.  
Next, the attention weight  is mapped into the action space by a sigmoid function to determine whether to add more edges for channel $m$. Initially when $\mathcal{E}=\emptyset$, $f_{addedge}(\mathcal{G},m)$  always outputs one, indicating that channel $m$ needs to be assigned with some beam nodes. 

Once $f_{addedge}(\mathcal{G},m)=1$, then each beam $b$ is evaluated by $f_{node}(\mathcal{G},m,b)=$ $softmax[f_{n}(m,\boldsymbol{n}_{\mathcal{E}}, \boldsymbol{g}_{\mathcal{E}})]$ to determine which beam to assign with channel $m$. 
Similarly, $f_{n}$ computes an attention weight for each potential edge $e_{m,b}$, and the softmax function calculates the possibility of adding each edge into $\mathcal{E}$.    
Furthermore, one condition is added in the output layer of $f_{n}$ to ensure that constraint of the maximum channel aggregation in (\ref{con_freq}) is satisfied. 
After adding a new edge, information propagation is applied for each beam to optimize its power. 

The edge generation towards channel  $m$ continues between two aforementioned steps until  $f_{addedge}(\mathcal{G},m)$ returns zero. 
Next, the algorithm proceeds  to $m+1$, and repeats the decision-making process until all $M$ channels are assigned.  
Finally, the beam-frequency graph is constructed with corresponding power allocations in the node feature, as summarized in Algorithm \ref{algo}. 

\subsubsection{Output distribution and training}
The generative graph model defines the probability distribution over all feasible graphs $P(\mathcal{G},\boldsymbol{n}_\mathcal{E},\boldsymbol{g}_\mathcal{E})$, i.e., the distribution of all beam-frequency allocations with corresponding downlink latency $\tau_s(\mathcal{G},\boldsymbol{n}_\mathcal{E},\boldsymbol{g}_\mathcal{E})$.  
Based on the above analysis, the probability for each beam $b$ to be assigned with channel $m$ can be given by  
$Pr(e_{m,b}\in\mathcal{E}) = \sum_{j=1}^{J_m} [ f_{node}(b|\mathcal{G}_j^m)\cdot \prod_{i=1}^{j} f_{addedge}(m|\mathcal{G}^m_i) ]$, 
where $J_m \in \{1,\cdots,B\}$ is the number of beam nodes associated to channel $m$, and $\mathcal{G}_j^m$ denotes the graph at the $j$-th iteration during the edge generation for channel $m$. 
Thus, the probability of a constructed graph can be given as 
$Pr(\mathcal{G}) = \prod_{m\in \mathcal{M}, b \in \mathcal{B}} Pr(e_{m,b}\in\mathcal{E})$, and the expected latency will be $ \mathbb{E}(\tau_s) = \sum_{\mathcal{G}} P(\mathcal{G},\boldsymbol{n}_\mathcal{E},\boldsymbol{g}_\mathcal{E})\tau_s(\mathcal{G},\boldsymbol{n}_\mathcal{E},\boldsymbol{g}_\mathcal{E})$.  
We consider both the decision-making functions $f_{addedge}$ and $f_{node}$  to employ neural networks with parameters of $\boldsymbol{\alpha},\boldsymbol{\eta}$, respectively, then the training of  graph neural network is to minimize the expected latency of the generated graph, i.e.,
$\min_{\boldsymbol{\alpha},\boldsymbol{\eta}} \mathbb{E}[\tau_s(f_{ae}(\boldsymbol{\alpha}), f_n(\boldsymbol{\eta}))]$.


\begin{algorithm}[h] 
	\caption{Beam-frequency allocation} \label{algo}
	\begin{algorithmic}
		\State \textbf{Initialize} $\mathcal{G}=\{ \mathcal{B}\cup\mathcal{M}, \mathcal{E}=\emptyset\}$, $\boldsymbol{n}_{b}$  and $\boldsymbol{n}_{m}$, $\forall b,m$;  \\   
		\textbf{For} $m$ in $[1, \cdots, M]$, \textbf{loop}\\ 
		\quad Update  $\boldsymbol{n}_{\mathcal{E}}$ and $\boldsymbol{g}_{\mathcal{E}}$, and compute $f_{addedge}(\mathcal{G},m)$ \\		
		\quad \textbf{If} $f_{addedge}(\mathcal{G},m)=1$ \\
		\quad \quad Compute  $b^*=\arg \max_b f_{node}(\mathcal{G},m,b)$\\
		\quad \quad Add $e_{m,b^{*}}$ to $\mathcal{E}$ \\ 
		\quad \textbf{Else}  break the loop\\ 
		\textbf{Return} $\mathcal{G}$, $\boldsymbol{n}_\mathcal{E},\boldsymbol{g}_\mathcal{E}$. 
	\end{algorithmic}
\end{algorithm} 
 
\vspace{-0.2cm}
 
\section{Simulation Results and Analysis}

In the simulation, we consider a Walker Delta constellation of $240$ LEO satellites at an altitude of $h$$=$$550$ km with the minimum elevation angle  $\theta_{min}$$=$$25^\circ$. 
The downlink frequency is centered at $12$ GHz with $M$$=$$8$ frequency channels, each with a bandwidth of $\beta$$=$$250$ MHz \cite{del2019technical}. 
Each satellite is equipped with $B$$=$$16$ spot beams with the maximum equivalent isotropic radiated power (EIRP) of $15$ dBW/MHz, the maximum antenna gain is $G^T_{max}$$=$$40$ dBi, one half of $3$-dB beamwidth is $\phi^h$$=$$1^\circ$, and the sidelobe edge angle is $\phi^{max}$$=$$20^\circ$. 
At the UT side, the maximum antenna gain is $G^R_{max}$$=$$35$ dBi, the half beamwidth is $\psi^e$$=$$1^\circ$, and the sidelobe edge angle is $\psi^{max}$$=$$40^\circ$. 
For performance comparison, two baselines are introduced. In the first approach of full frequency reuse, each beam employs all $M$ frequency bands with even transmit power allocation. In the second approach, each beam selects only one frequency band, concentrating all transmit power on the single channel band.  

\begin{figure}[!t]\vspace{-0.4 cm} 
	\begin{center}  
		\includegraphics[width=7.8cm]{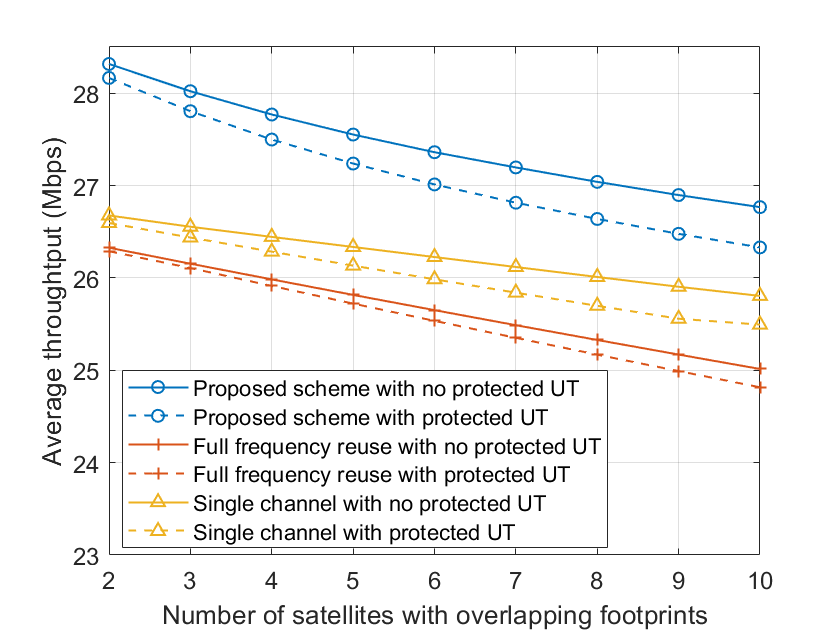}  \vspace{-0.2 cm}
		\caption{\label{rate_neighSate}\small The average downlink throughput decreases as the number of neighboring satellites with overlapped footprints increases.}  
	\end{center}\vspace{-0.7cm}  
\end{figure}

Fig. \ref{rate_neighSate} shows that the average downlink throughput decreases as the number of neighboring satellites with overlapping footprints $|\mathcal{L}_s|$ increases. In LEO constellations, the value of  $|\mathcal{L}_s|$ varies as satellite $s$ moves along its orbital path. 
Inter-orbit distances are largest  near the equator, where overlapping coverage typically involves only two satellites from adjacent orbits. Conversely, satellite density is significantly higher near polar regions, resulting in increased CCI and reduced throughput per satellite. 
Compared to baseline methods, our proposed approach achieves better performance, by intelligently assigning frequency bands to each beam to minimize the CCI within the constellation and optimizing power allocation to enhance downlink data rates. 
Additionally, the single-channel scheme outperforms full frequency reuse due to its flexibility in selecting frequency channels selection to avoid CCI when beams are oriented in similar directions. 
Furthermore, to evaluate the impact of EPFD-protect UTs on LEO downlink communications, we consider a radio astronomy telescope randomly located within the footprint of a LEO satellite, with $\kappa$$=$$-$$160$ dBW/m$^2$ per $100$ MHz, and its transmission frequency overlaps with one of satellite communication channels.   
In this scenario, all methods experience performance degradation due to the necessary reduction in transmit power to prevent CCI with the protected UT. 
Notably,  as shown in Fig. \ref{rate_neighSate},  LEO satellite throughput is no more than $30$ Mbps due to the assumption of servicing one UT per beam. Future work will explore simultaneous service to multiple UTs per beam, which is expected to enhance throughput to the Gbps level. 
  
\begin{figure}[!t]\vspace{-0.3 cm} 
	\begin{center}  
		\includegraphics[width=7.6cm]{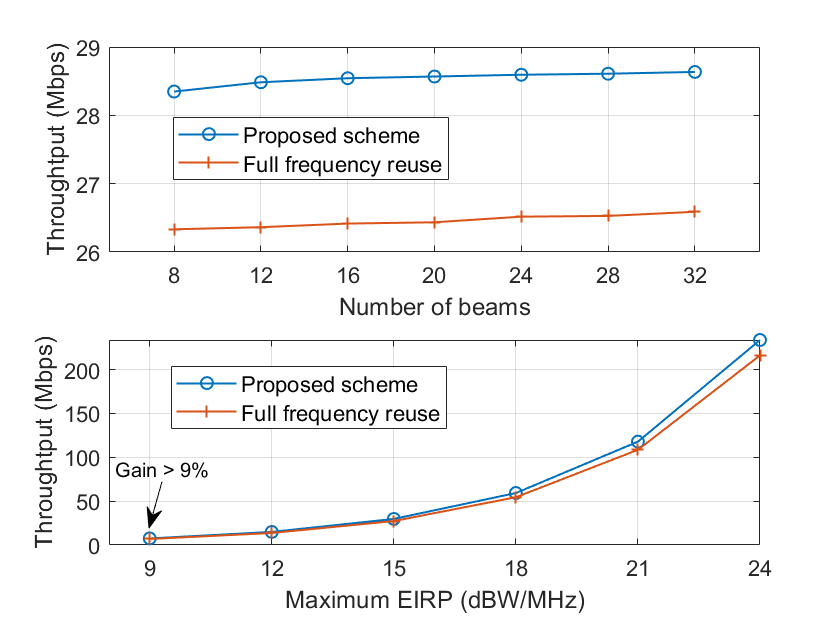} \vspace{-0.25 cm} 
		\caption{\label{rate_power_beam}\small The average downlink throughput increases with the number of beams and the transmit power. }  
	\end{center}\vspace{-0.45cm}  
\end{figure}

Fig. \ref{rate_power_beam} illustrates that average downlink throughput increases with both the number of beams and the transmit power. 
In particular, when the number of overlapping satellites is two, the LEO system is more power-constrained than beam-limited. 
As shown in Fig. \ref{rate_power_beam}, doubling the number of beams results in only a marginal increase in the average throughput,  
because given a fixed transmit power, the downlink power will be divided among more beams, limiting improvements in the average communication rate.  
On the other hand, given interference from two overlapping satellites is manageable, doubling the transmit power nearly double the downlink transmission rate. 
Additionally, the proposed resource allocation approach improves the average throughput per satellite by more than $9\%$, compared to the full frequency reuse scheme.  

\begin{figure}[!t]\vspace{-0.25 cm} 
	\begin{center}  
		\includegraphics[width=7.6cm]{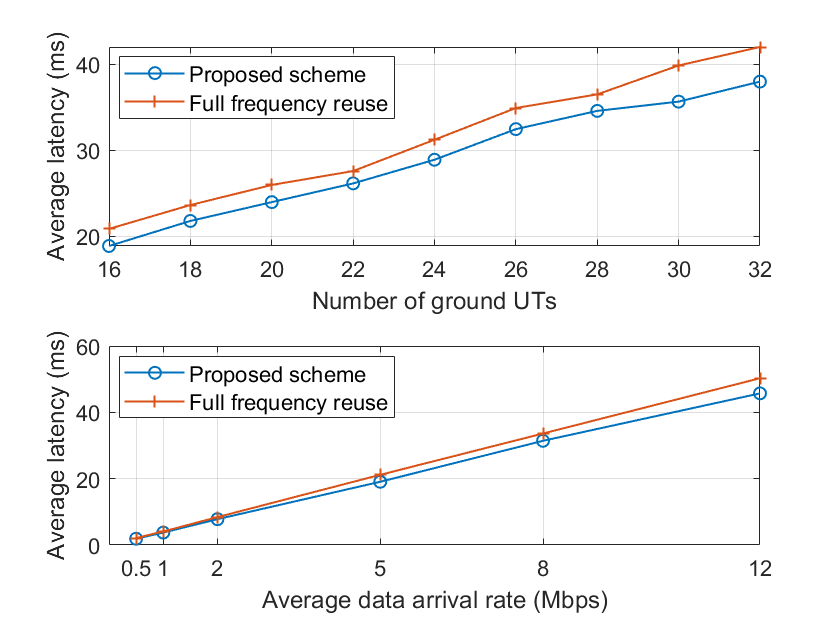} \vspace{-0.25 cm} 
		\caption{\label{latency}\small Average latency per UT increases given a higher number of ground UTs and an increased rate of data packet arrivals per second. }  
	\end{center}\vspace{-0.7cm}  
\end{figure}

Fig. \ref{latency} shows that the average latency per UT increases if the total number of ground UTs served by the considered LEO satellite increases or the average data rate of downlink package arrivals becomes higher.  
Basically, the latency increase follows a linearly pattern, with the proposed methods outperforming the full frequency reuse scheme by approximately $10\%$. 
Furthermore,  similar to throughput, multi-user transmission can further reduce the latency, especially in scenarios with high-density users, which will be explored in our future work.


\section{Conclusion}
 
In this paper, we have proposed a novel  framework  for optimizing resource allocation and beam scheduling in multi-beam LEO communication systems. 
To support efficient transmissions, a hybrid beam pattern has been employed to enhance the quality of service for downlink communications. 
A dynamic CCI control mechanism has been developed to mitigate inter-beam interference within LEO constellations and reduce cross-system interference affecting protected users from other communication networks. 
To address the problem of beam-UT-frequency allocation with power optimization,  a graph generation algorithm has been proposed with low computational complexity to enable a real-time network operation and minimize the downlink latency.  
Simulation results have demonstrated that the proposed approach outperforms the baseline methods  of full frequency reuse and single-channel transmission.

\bibliographystyle{IEEEtran}
\bibliography{references}

\end{document}